%
\let\useblackboard=\iftrue
%
%
\newfam\black
\input harvmac.tex
\useblackboard
\message{If you do not have msbm (blackboard bold) fonts,}
\message{change the option at the top of the tex file.}
\font\blackboard=msbm10 scaled \magstep1
\font\blackboards=msbm7
\font\blackboardss=msbm5
\textfont\black=\blackboard
\scriptfont\black=\blackboards
\scriptscriptfont\black=\blackboardss

\else

\fi
%
\def\yboxit#1#2{\vbox{\hrule height #1 \hbox{\vrule width #1
\vbox{#2}\vrule width #1 }\hrule height #1 }}
\def\fillbox#1{\hbox to #1{\vbox to #1{\vfil}\hfil}}
\def\ybox{{\lower 1.3pt \yboxit{0.4pt}{\fillbox{8pt}}\hskip-0.2pt}}

\def\comments#1{}

\def\Tr{{{\rm Tr\  }}}
\def\tr{{\rm tr\ }}

\def\Im{{\rm Im\hskip0.1em}}

\def\ket#1{|#1\rangle}
\def\vev#1{\langle{#1}\rangle}

\def\CM{{\cal M}}
\def\CN{{\cal N}}

\def\sqap{\sqrt{\alpha'}}

\def\II{\relax{I\kern-.10em I}}

\def\IZ{\relax\ifmmode\mathchoice
{\hbox{\cmss Z\kern-.4em Z}}{\hbox{\cmss Z\kern-.4em Z}}
{\lower.9pt\hbox{\cmsss Z\kern-.4em Z}}
{\lower1.2pt\hbox{\cmsss Z\kern-.4em Z}}\else{\cmss Z\kern-.4em
Z}\fi}
\def\IB{\relax{\rm I\kern-.18em B}}
\def\IC{{\relax\hbox{$\inbar\kern-.3em{\rm C}$}}}
\def\ID{\relax{\rm I\kern-.18em D}}
\def\IE{\relax{\rm I\kern-.18em E}}
\def\IF{\relax{\rm I\kern-.18em F}}
\def\IG{\relax\hbox{$\inbar\kern-.3em{\rm G}$}}
\def\IGa{\relax\hbox{${\rm I}\kern-.18em\Gamma$}}
\def\IH{\relax{\rm I\kern-.18em H}}
\def\II{\relax{\rm I\kern-.18em I}}
\def\IK{\relax{\rm I\kern-.18em K}}
\def\IP{\relax{\rm I\kern-.18em P}}

\def\inbar{\,\vrule height1.5ex width.4pt depth0pt}

\font\cmss=cmss10 \font\cmsss=cmss10 at 7pt
\def\IR{\relax{\rm I\kern-.18em R}}

\def\Tr{\rm Tr}
\def\BR{\IR}
\def\BZ{\IZ}
\def\BP{\IP}
\def\BR{\IR}
\def\BC{\IC}

\lref\dkps{
M. R. Douglas, D. Kabat, P. Pouliot and S. H. Shenker,
``D-Brane and Short Distances in String Theory,''
hep-th/9608024.}
\lref\bfss{T. Banks, W. Fischler, S. Shenker, S. Susskind,
``M Theory as a Matrix Model: a Conjecture,''
hep-th/9610043.}
\lref\polD{J. Polchinski, ``Dirichlet Branes and
Ramond-Ramond Charges,''
Phys. Rev. Lett.75 (1995) 4724, hep-th/9510017.}
\lref\bachas{
C. Bachas and M. Porrati,
```Pair Creation of Open Strings in an Electric
Field,'' Phys. Lett. 296B (1992)
77, hep-th/9209032; C. Bachas, ``D-brane Dynamics,'' Phys. Lett. B374
(1996) 37, hep-th/9511043.}
\lref\bachastwo{C. Bachas and E. Kiritsis, ``F(4) Terms in N=4 String
Vacua'', hep-th/9611205.}
\lref\douglasli{M. R. Douglas and
M. Li, ``D-Brane Realization of $N=2$ Super Yang-Mills
Theory in Four-Dimensions,'' hep-th/9604041.}
\lref\dm{M. Douglas and G. Moore, ``D-Branes, Quivers, and ALE Instantons,''
hep-th/9603167.}
\lref\curved{
For CFT discussion of $D$-brane on $K3$ or Calabi-Yau
manifolds, see:
H. Ooguri, Y. Oz and Z. Yin,
``D-Branes on Calabi-Yau Spaces and Their Mirrors,''
Nucl. Phys. B477 (1996) 407, hep-th/9606112.}
\lref\et{T. Eguchi and A. Taormina,
``Character Formulas for the $N=4$ Superconformal
Algebra,''
Phys. Lett. 200B (1988) 314;
``On the Unitary Representations of $N=2$ and $N=4$
Superconformal Algebras,''
Phys. Lett. 210B (1988) 125.}
\lref\egs{M. R. Douglas, ``Enhanced Gauge Symmetry in M(atrix)
Model,'' hep-th/9612126.}
\lref\kp{D. Kabat and P. Pouliot, 
``A Comment on Zero-Brane Quantum Mechanics,'' Phys. Rev. Lett.
77 (1996) 1004, hep-th/9603127.}
\lref\dfs{U.~H.~Danielsson, G.~Ferretti, and B.~Sundborg, ``D-particle
Dynamics and Bound States,'' hep-th/9603081.}
\lref\cpt{W. Taylor, ``D-brane Field Theory on Compact Spaces,''
hep-th/9611042; O. Ganor, S. Rangoolam and W. Taylor, ``Branes, Fluxes
and Duality in (M)atrix Theory,'' hep-th/9611202.}
\Title{
\vbox{\baselineskip12pt\hbox{hep-th/9702203}\hbox{RU-97-09}
\hbox{UCB-PTH-97/07}\hbox{LBNL-40003}
}}
{\vbox{
\centerline{Issues in M(atrix) Theory Compactification} }}
\centerline{Michael R. Douglas$^1$, Hirosi Ooguri$^{2,3}$
and Stephen H. Shenker$^1$}
\medskip
\centerline{$^1$ Department of Physics and Astronomy}
\centerline{Rutgers University}
\centerline{Piscataway, NJ 08855--0849}
\centerline{\tt mrd, shenker @physics.rutgers.edu}
\smallskip
\centerline{$^2$ Department of Physics}
\centerline{University of California at Berkeley}
\centerline{Berkeley, CA 94720--7300}

\smallskip
\centerline{$^3$ Physics Division,
Mail Stop 50A--5101}
\centerline{
E. O. Lawrence Berkeley National Laboratory}
\centerline{Berkeley, CA 94720}
\centerline{\tt ooguri@physics.berkeley.edu}
\bigskip
\centerline{\bf Abstract}
We discuss issues concerning M(atrix) theory compactifications on curved
spaces.  We argue from the
form of the graviton propagator on curved space that excited string
states do not decouple from the annulus D0-brane
$v^4$ amplitude, unlike the flat
space case.  This argument shows that a large class of quantum
mechanical systems with a finite number of degrees of freedom cannot
reproduce supergravity answers.  We discuss the specific example of an
ALE space and suggest sources of possible higher derivative terms that
might help reproduce supergravity results.

%
\noindent

\Date{February 1997}
\newsec{Introduction}
The conjectured description of M theory as matrix quantum mechanics
\bfss\ implies
a substantial reduction in the apparent degrees of freedom necessary
to describe M theory.  A  first reduction occurs
because various different p-branes emerge as composites of
of the $N$ D0-branes described by the $U(N)$ quantum mechanics as
$N \rightarrow \infty$.
The second reduction occurs because all excitations
of the stretched string states connecting the D0-branes decouple from
the leading low velocity dynamics, allowing classical
supergravity interactions to emerge as the result of integrating
out a finite number of quantum mechanical degrees of freedom.
This phenomenon was found in the
$v^4$ ($v$ is the D0-brane velocity) leading weak coupling (annulus)
amplitude in \dkps.  As explained very clearly in \bachastwo\ it follows
from the vanishing contribution of long (non-BPS) $N=4$ supersymmetry
multiplets to this amplitude.

The scenario for proof outlined in
\bfss\ requires that i) all relevant velocities go to zero as
$N \rightarrow \infty$ and ii) that there are no corrections beyond one
loop in string theory to the leading low velocity dynamics.  With these
two conditions the decoupling of excited string states would extend to
arbitrarily strong coupling.  So the D0-brane $U(N)$ quantum mechanics which
describes only the lowest unexcited stretched string state dynamics
would provide an accurate strong coupling description.  More generally,
what is required is that the excited string states decouple from the
full leading low velocity dynamics.

A priori, the excited state decoupling would be expected only for
maximal (e.g. $N=4$, $d=4$) spacetime supersymmetry.
More generally, the leading weak coupling $O(v^4)$
interaction crosses over
from supergravity at distances $l >> \sqap$ to a sum over exchanges
of all closed strings, equivalent to a quantum mechanical open
string amplitude where all excited open strings contribute.
At distances $l >> l_P^{11}$ (the eleven dimensional Planck length),
this is essentially a one-loop amplitude.

In section 2, we show that this crossover is non-trivial in string theory
whenever there is nonvanishing curvature in the compactified space.
In other words, the form of the $O(v^4)$ interaction predicted by
supergravity never agrees with the truncation of the
one-loop open string amplitude result to
a finite number of quantum mechanical degrees of freedom.

In section 3 we review the theory of D$0$-branes at weak string
coupling on the orbifold
$\BC^2/\BZ_2$ and its smooth resolution (Eguchi-Hanson space),
developed in \dm.
For present purposes, the main point is that
this is a quantum mechanics with a finite number of degrees of freedom,
to which the argument of section 2 applies.
We compare the quantum mechanics
and supergravity results and point out another mismatch with
supergravity -- the mass of
stretched open strings is apparently not proportional to their length --
which may be resolved by further computation within the framework of \dm.

In section 4, we outline a version of this definition which
is well motivated at strong coupling (extending results of \egs), and
we explain ways in which it might evade the theorem of section 2.

We discuss some implications of these results in section 5.

\newsec{Annulus}

We begin by computing the scattering of two $D0$-branes on
$R^6 \times K3$. We consider the case when
they are fixed at points in $K3$, but move
on $R^6$ with relative velocity $v$ and impact parameter
$b$. By combining
results of \polD, \bachas, \douglasli\ and \dkps, we
find that their scattering amplitude is given by
\eqn\scat{
{\cal A} = \int {dt \over t}
e^{-t b^2/2 \pi \alpha'} {1 \over \eta(t)^4}
\left( {\theta_{11}'(0|t) \over \theta_{11}(\epsilon t|t)} \right)
\times Z(t,\epsilon)}
where
\eqn\fermion{
\eqalign{ Z(t, \epsilon) = &
{\rm Tr}_{NS}\left( q^{L_0-1/4}\right)
{\theta_{00}(0|t) \theta_{00}(\epsilon t|t) \over 2 \eta(t)^2}
-{\rm Tr}_{NS}\left( (-1)^F q^{L_0-1/4}\right)
{\theta_{10}(0|t) \theta_{10}(\epsilon t|t) \over 2 \eta(t)^2} - \cr
& -{\rm Tr}_{R}\left(q^{L_0}\right)
{\theta_{01}(0|t) \theta_{01}(\epsilon t|t) \over 2 \eta(t)^2} \cr}}
The parameter $\epsilon$ is related to the velocity
$v$ as $\pi \epsilon = {\rm arctanh}(v)$.
The traces
\eqn\traces{ {\rm Tr}_{NS} \left( q^{L_0-1/4}\right),
   ~~{\rm Tr}_{NS} \left( (-1)^F q^{L_0-1/4}\right),
   ~~{\rm Tr}_{R} \left( q^{L_0}\right), }
are the partition function of
the open string on $K3$. The CFT of
closed string on $K3$ has two copies of the
$N=4$ superconformal symmetry on the worldsheet.
Since the D0-brane boundary condition preserves
1/2 of them \curved, at least one
set of the $N=4$ superconformal algebra acts
on the open string Hilbert spaces. We can therefore
expand the open string partition functions
in terms of the characters of
the $N=4$ algebra with $\hat{c} = 2$ studied in
\et .
There are three types of representations of the $N=4$ algebra.
Two of them have non-zero values of Witten index and
are called the massless representations. Their conformal
weights are $0$ and $1/2$ and their characters
$\chi_0$ and $\chi_{1/2}$ obey the following simple relation,
\eqn\massless{
   \chi_0^{(NS)} + 2 \chi_{1/2}^{(NS)} = {q^{-1/8} \over \eta}
  \left( {\theta_{00} \over \eta} \right)^2 }
In particular, the Witten indices of the two representations
cancel in this combination after performing the spectral
flow to the R-sector.
The third type is called a massive representation; it
has no Witten index and it exists for any conformal weight
$h>0$. The character of massive representation is
\eqn\massless{
   \chi_h^{(NS)} = {q^{h-1/8} \over \eta}
  \left( {\theta_{00} \over \eta} \right)^2}

Now it is easy to see that the open string Hilbert space
in question has no Witten index. This is because the string
stretched between the two points on $K3$ has a non-zero
energy proportional to the geodesic distance
between them. Although
this is a semi-classical statement valid for distances
much larger than the string length $l_s \sim \sqrt{\alpha'}$, the fact that
the Witten index vanishes is rigorous. Therefore the partition
functions \traces\ of the open string should be given by
\eqn\open{
{\rm Tr}_{NS} \left( q^{L_0 - 1/4} \right)
 = {g(t) q^{-1/8} \over \eta}\left( {\theta_{00} \over \eta}
\right)^2 }
where $g(t)$ encodes multiplicities of the $N=4$ algebra
representations,
\eqn\spectrum{ g(t) = \sum_i q^{h_i}  . }
Substituting this into \fermion, we find
\eqn\integrand{
  Z(t,\epsilon) = {g(t) q^{-1/8} \over 2 \eta(t)^5}
 \Big[ \theta_{00}(0|t)^3 \theta_{00}(\epsilon t|t)
     - \theta_{10}(0|t)^3 \theta_{10}(\epsilon t|t)
  - \theta_{01}(0|t)^3 \theta_{01}(\epsilon t|t)  \Big] .}

When $v$ is small, we can expand
\scat\ in powers of $\epsilon \sim v/\pi$. The $v^2$ term
vanishes as in the case of the flat space \bachas\ since
$$ Z(t, \epsilon) = O(v^4) . $$
Using the result of \dkps,
the coefficient of the $v^4$-term can be expressed
as
\eqn\vfour{
  {\cal A}_{v^4} = \int {dt \over t}
e^{-t b^2/2 \pi \alpha'} g(t) \prod_{n=1}^\infty (1-q^n)^3. }

The question is whether only the lightest stretched string state
contributes to $g(t)$.  We now argue that this cannot be the case.
To understand why, we observe that on general string theory grounds
${\cal A}_{v^4}$ approaches the result of massless closed
string (supergraviton) exchange when the distances between the
D0-branes are much larger than $l_s$.
In this domain we have
\eqn\false{
 {\cal A}_{v^4} \rightarrow \int {dt \over t}
 \left( {t \over \alpha'} \right)^2
 e^{-t b^2/2 \pi \alpha'}
           \left( e^{- {\alpha' \over t} \Delta_{K3}} \right)_{x,y}, }
where $\Delta_{K3}$ is the Laplacian on $K3$ and  $x,y$
are points on $K3$ where the D0-branes are located.

If \vfour\ and \false\ are to agree for all $b>>l_s$, we would require
$$ g(t) = {
(\alpha'/t)^2 \over \prod_{n=1}^\infty (1-q^n)^3}
   \left( e^{- {\alpha' \over t} \Delta_{K3}} \right)_{x,y}. $$
for all $t<<1$.
This is not possible. To see this, let us expand the right-hand side
for large $t$ (small $q=e^{-t}$). It can be done by using
the adiabatic expansion of the heat kernel $e^{- {\alpha' \over t}
\Delta} $, and we find
\eqn\impossible{
  g(t) = {1 \over \prod_{n=1}^\infty (1-q^n)^3}
   e^{-{t \over \alpha'} \sigma^2(x,y)}
 \left[ 1 + \sum_{k=1}^\infty a_k(x,y) \left( {\alpha' \over
t} \right)^k \right], }
where $\sigma(x,y)$ is the geodesic distance between $x$ and $y$,
and $a_k(x,y)$ can be expressed in terms of the curvature of $K3$,
$\sigma(x,y)$, and their derivatives. We know that some of these
coefficients are non-zero; in fact the first term is the Euler density.
This expansion is valid for $t>>\sigma^2/l^2_C$ where $l_C$
is the characteristic curvature length.  So  for $l_s^2<<\sigma^2<<l^2_C$
there is a large region $1>>t>>\sigma^2/l^2_C$ where \impossible\
is valid.
On the other hand, we know $g(t)$ must have an expansion of
the form \spectrum.  No finite number of states, or discrete infinity
of states whose gaps are not string scale or smaller, can reproduce the
form \impossible\ in the required range of $t$. On other hand
the excited open string states can produce such effects and therefore must
contribute.  The unexcited multiple winding states important for describing
D0-brane dynamics in toroidal compactifications \refs{\bfss,\cpt},
or multiply wound extremal geodesic open strings,
have gaps $\sim l_C^2$ and
do not affect the above conclusions.

A similar argument to the above demonstrates that excited open string
states must contribute in Calabi-Yau  compactifications
with spacetime $N=1$ supersymmetry
as well.  Again, the crucial point is that in general there is nonzero
curvature in such compactifications,
hence nontrivial power law corrections in \impossible\
which cannot be reproduced by the unexcited open string states.

Although the result is generic for the quantum mechanics
of weakly coupled
open strings, there are a number of implicit assumptions which might
be violated in more general contexts, such as a quantum mechanics
of M theory compactification.
Perhaps the most serious is the decoupling between
$R^6$ and $K3$ world-sheet degrees of freedom.
This led directly to the factorized nature of the amplitude
\open, and the simple $b$ dependence in \vfour.
In terms of quantum mechanics, it restricts the masses
of states to depend on the parameters in $\BR^6$ as
\eqn\wmass{
m_W^2 \sim b^2 + f(x,y)
}
with no explicit $v$ dependence.
On the other hand, if we allow general dependence on $b$ and $v$, the
supergravity result could be reproduced in many ways.

\newsec{Interactions of D$0$-branes on ALE space}

In \dm\ it was found that $N$ D$0$-branes on $\BC^2/\BZ_2$ are described
by the dimensional reduction of $\CN=1$, $d=6$
$SU(N)\times SU(N) \times U(1)\times U(1)$ gauge
theory with two hypermultiplets in the $(N,\bar N)_{(2,0)}$.  The
parameters $\zeta$
which blow this up to the Eguchi-Hanson space $\CM_\zeta$ are simply
the three Fayet-Iliopoulos terms for the non-trivial $U(1)$.

The strategy for defining D$0$-branes on an orbifold
is identical in string theory and in M
theory, but we briefly review it in the latter framework.
We start with the maximally supersymmetric $U(2N)$ quantum mechanics,
and make a projection commuting with half of the supersymmetry,
\eqn\orbproj{\eqalign{
\omega X &= \gamma X \gamma^{-1} \cr
\omega^{-1} \bar X &= \gamma X \gamma^{-1} \cr
A &= \gamma A \gamma^{-1}
}}
with $\omega=-1$ and $\gamma=\sigma_3\otimes{\bf 1}_N$.
Each boson has a partner fermion with the same projection.
The bosonic matter is
\eqn\bmat{
X = \left(\matrix{0&b_{01}\cr b_{10}&0}\right) \qquad
\bar X = \left(\matrix{0&\bar b_{10}\cr \bar b_{01}&0}\right).
}

The resulting Lagrangian is determined by the choice of
gauge group and matter representation, if we assume the
the matter Lagrangian is free before gauging.  We will make this
assumption, but discuss it further below.

We next review the identification of the Higgs
branch of the moduli space with $\CM_\zeta$.
The analysis can be done for $N=1$; the complete Higgs branch for $N>1$ is
the obvious symmetric product $\CM_\zeta^N/S_N$.
This is the hyperk\"ahler quotient constructed by Kronheimer, defined by
the three moment map (D-term) constraints
\eqn\Dterms{\eqalign{
b_{01}\bar b_{01} - b_{10}\bar b_{10} &= \zeta_C \cr
|b_{01}|^2 - |b_{10}|^2 - |\bar b_{01}|^2 + |\bar b_{10}|^2 &= \zeta_R
}}
and the $U(1)$ gauge quotient.
In the $\BZ_2$ (Eguchi-Hanson) case, there is an $SU(2)$ symmetry under
rotations of the vector $\vec\zeta$, allowing us to take $\zeta_C=0$
and $\zeta_R>0$ without loss of generality.  We do so below.

The $\BP^1$ produced by blowup is then
\eqn\twosphere{b_{10}=\bar b_{01}=0.}
Taking
\eqn\twospherecoord{ z_1\equiv b_{01} \qquad z_2\equiv\bar b_{10}, }
the remaining constraint and quotient become the usual K\"ahler quotient
construction of the Fubini-Study metric on $\BP^1$.
Even more simply,
we can generically gauge $\Im\bar b_{01}$ to zero, and the constraint
becomes the usual $\sum x_i^2 = \zeta$ defining $S^2\in \BR^3$.

This quantum mechanics certainly falls under the hypotheses of the
result in section two, and we conclude that the one-loop $O(v^4)$
interaction energy between the two D$0$-branes cannot reproduce the
subleading corrections in \impossible.

In fact the situation is worse -- it does not reproduce the
leading term.  To see this, we compute the masses of the W bosons.
On the Higgs branch, $U(2)\times U(2)$ is broken to
$U(1)\times U(1)$, and thus these fall into $6$ massive multiplets
(hyper + vector) of $\CN=2$.  Each contains a massive vector boson
whose mass matrix is $\Tr [A^i,\vev{X}][A^j,\vev{X}]$,
as in any Yang-Mills theory; all states in the multiplet have this mass.

This mass matrix is a truncation of that in
the D-brane theory before applying the projection \orbproj; furthermore
gauge bosons with different eigenvalues under the projection operator
$\gamma A \gamma^{-1}$ do not mix; therefore
the mass of a stretched string is proportional to its length in the
configuration space of the unprojected theory.
Finally,
since the vevs $\vev{X}$ are a linear subspace of those possible in
the unprojected theory, this length is the same as the distance
$\sqrt{\tr (X-X')^2}$ in the configuration space of the projected theory.

The conclusion is that
the mass of a stretched string
is proportional to its length in the larger configuration space.
For the special case of two D$0$-branes located on the two-sphere
\twosphere, we can use symmetry to set $z_2=z_2'=0$ leaving
two positions $z_1$ and $z_1'$ with $|z_1|^2=|z_1'|^2=\zeta_R$.
Then there are two massive multiplets with mass $|z_1-z_1'|$ (the
distance between the two branes),
two with mass $2\sqrt{\zeta_R}$ (the distance between a brane and its image),
and two with mass
$|z_1+z_1'|$ (a brane and the other's image).
In other words, the strings take the
shortest path between D-branes (and their images), which passes inside
the two-sphere, not on it.  A shortest geodesic distance would of course
be $\theta = \Im\log z_1'/z_1$.

Thus we find the leading short distance behavior
$\exp -|z_1-z_1'|^2 t/\alpha'$ for the function $g(t)$ in \vfour,
in contradiction with the leading term $\exp -\sigma^2 t$
in \impossible.
Now this result is not in obvious contradiction with physical expectations
for sub-stringy physics at weak string coupling.
In principle, exchange of massive closed
strings could combine to this answer.
However, it is suspicious.

In fact, the result depends on an assumption which has
not been proven: that the kinetic term in the world-volume gauge theory
is the trivial $\sum |b_i|^2$.
This is the simplest guess at a metric which on grounds of supersymmetry must
be hyperk\"ahler and admit a $SU(N)\times SU(N) \times U(1)$ isometry,
but it has not been proven that
it is the unique candidate.
{}From world-sheet considerations along the lines of \dm,
the metric will be flat at $\zeta=0$ with corrections
computable in weak string coupling.
This computation and the question of whether it will modify
this result are presently under study.

\newsec{D$0$-branes on ALE space at strong coupling}

In studying the same system at strong string coupling, we are forced to
rely more on consistency arguments.

One natural idea is to start with gauge theory with a curved target
space and dimensionally reduce it.  This is interesting but will almost
certainly produce a singular theory in the orbifold limit and as such
is not likely to be the correct definition for small blow-up parameter.

Another natural approach which works well in the case of toroidal
compactification is to introduce images under the space group.
In the case of ALE this is exactly what we did above in defining
$\BC^2/\BZ_2$.  Clearly the correct supergravity interactions are obtained
at one loop -- they are the sum of image contributions.

The second step of adding the FI terms to produce the blowup
(as proposed in \egs) is motivated by the observation
that with this amount of supersymmetry
there is nothing else we can do that changes the topology of moduli space.
However, higher order corrections to the Lagrangian are
less restricted.\foot{
The explicit claims of \egs\ did not depend on these.
}
In the orbifold limit, the lack of
any scale (other than the overall coupling $l_{P}^{11}$) makes it very
plausible that such corrections vanish.

To work around the better controlled orbifold limit, we can consider the
blow-up to be accomplished by adding a condensate of particles in the
multiplet associated with the blow-up mode.\foot{
This idea and the related idea discussed in the conclusions
arose in conversation with Tom Banks.}
These can be identified with bound states of D$0$-branes, as discussed
in \egs.  Although in the IMF
it is not possible to make a spatially independent condensate this way,
the minimal accessible longitudinal momentum $p_-$ goes to zero
(as $1/N$) in the large $N$ limit of \bfss, and the result is effectively
a constant blow-up for our purposes.
For $\BC^2/\BZ_2$, the minimal bound state which respects the $U(1)$
manifest in the geometric description is a bound state in the $N=1$
model.

The condensate is most simply described by introducing second quantized
operators which relate Hilbert spaces of different $N$.  A creation operator
$B^+$ for the bound state we mentioned would act on the $N$ particle
Hilbert space and produce a state in the $N+1$ particle Hilbert space.
In the limit where the $N$ preexisting particles are far from the fixed
point, the operation is simply tensor product.  We will not attempt a
general definition here (which probably requires knowing the bound
state wave function) but make the assumption that correlations between
different bound state (and other) particles can be neglected for our
purposes.  Then the wave function can be taken as a tensor product, and
expectation values will add in a simple way.  We then define an
annihilation operator $B$ as its adjoint.  The condensate is then
\eqn\condense{
\ket{\zeta} = e^{\zeta B^+} \ket{0}.
}

Given this definition, the bound state wave function,
and the assumption that correlations can be neglected,
it is straightforward in principle to deduce the Lagrangian describing
D$0$-branes on the blown-up orbifold.  The Fayet-Iliopoulos term must
come from evaluating
the potential $\sum_{i<j} \tr [X^i,X^j]^2$,
with a vacuum expectation value for the off-diagonal components $[X^i,X^j]$.
It would be interesting to verify this and
it may be possible to find a topological quantity containing this
expectation value by writing a trace projected to the BPS states,
something like $\Tr (-1)^F F [X^i,X^j] J_R^a$
where $J_R^a$ is the $SU(2)_R$ generator.

This procedure can generate fairly general higher order corrections
to the Lagrangian, with coefficients analytic in $\zeta$.
As at weak string coupling, the first question of interest is whether
corrections to the metric on moduli space can produce an effective
Lagrangian for which stretched strings have mass proportional to geodesic
length.  In the present case, we have strong physical reasons to expect this,
and the computation we have outlined will provide a significant test of this
proposal.

Let us assume that this works, and ask whether the resulting theory
can evade the result of section 2.
Since the condensate is constant in the transverse dimensions,
we expect a weaker version of the condition \wmass\ to hold,
\eqn\wmasstwo{
m_i^2 \sim b^2 + f_i(x,y,v).
}
However there is no reason for the theory to satisfy the stronger
world-sheet decoupling condition described there.

One can imagine one loop diagrams which could
reproduce higher order terms in the expansion \impossible.
A term $a_k/t^k$ must contain $k$ fewer
propagators for the states of mass squared $m^2=b^2+\sigma^2$, to make it
less singular.
Another necessary ingredient is that any singularities of the
coefficients $a_k$ in the orbifold limit must come from integrating out
states (since the orbifold limit is non-singular).  This suggests that
states of mass squared $m^2=b^2+\zeta$ must be present in the model,
and indeed they are.

The tentative conclusion is that this model might reproduce the
supergravity interaction, and work on testing this continues.

\newsec{Discussion}
In section 2 we have argued that no truncation to a finite
number of open string degrees of freedom can reproduce graviton exchange
in the annulus when the compactification has nonzero curvature.
This means that the quantum mechanics obtained by truncating to the
unexcited open string state will also fail to reproduce gravity at one
loop. 
The ``mild'' infinity of wrapped open strings introduced in
toroidal compactification \refs{\bfss , \cpt} also do not affect the result.
In principle, they might bring in the need for a cutoff and renormalization,
which could complicate the discussion.  It is interesting that the explicit
example in section 3, regarded as a 3-brane theory, is a finite quantum theory
(as was noted in \dm).  By the general connection between open string UV and  
closed string IR limits, this might be expected as a general feature of  
theories with no closed string IR divergences, e.g. with more than four  
non-compact dimensions.

Higher order loop corrections
in the string coupling $g_s$ are, by the scaling discussed in
\refs{\dfs, \kp, \dkps}, an expansion in $(l_P^{11}/r)^3$ where
$l_P^{11}$ is the eleven dimensional planck length and $r$ is a
characteristic separation length ($\sim b, \sigma$) of the D0-branes.
As long as the curvature length $l_C$ is large compared to $l_P^{11}$
these effects will not affect the arguments in section 2.

The argument clearly applies to the quantum mechanical model
discussed in section 3 and in fact to a wide class of quantum mechanical
models.  The main assumptions are that there be a finite
number of degrees of freedom, and restrictions on the higher derivative
terms, particularly couplings to the velocity.

What kind of matrix model description might work for such compactifications?
If the relevant velocities remain low one could recover the correct
supergravity by adding explicit higher derivative terms to the quantum
mechanics.
A rather simplistic example would be to include the explicit
$v^4$ term resulting from integrating out the excited open string
states in the annulus.
By itself this is probably not suitable (it would be singular in the
orbifold limit), but some combination of explicit and induced
interactions may well work, and the expansion in the blowup $\zeta$
described in section 4 might provide a theory of this type.

Large $N$ effects  might provide another way for quantum mechanics
to reproduce supergravity.  If the matrix model description is correct
in flat space it describes gravitons as bound states of D0-branes,
so we expect a compactification could be represented as some kind of condensate
of D0-branes.
In section 4 we discussed such a condensate, and suggested that its
effects could be summarized in an effective Lagrangian for a
finite degree of freedom system.
This is not logically necessary and it is also possible that
the only probes which have correct M theory physics are bound states
of large numbers of D0-branes in the large $N$ limit.
Perhaps the extended nature of the bound states enters crucially in this
physics.

In any event it does seem that any matrix model description of
M theory on curved
spaces will be rather more intricate than the flat space
description of \bfss.

\medskip

\centerline{\bf Acknowledgments}
We would like to thank Tom Banks, Greg Moore, Hiraku Nakajima,
Nathan Seiberg, and Lenny Susskind
for valuable discussions and correspondance.
H.O. thanks the High Energy Theory Group
of Rutgers University, where a part of this work was done, for
hospitality. H.O. is supported in part by NSF grant PHY-951497 and DOE
grant DE-AC03-76SF00098.  The work of M.R.D. and S.H.S. was supported in part 
by grant  No. DE-FG02-96ER40959.

\listrefs

\end